\documentclass[twocolumn, amsmath, twocolappendix]{aastex631}
\usepackage{amsmath}

\begin{document}

\title{Prospects for Observing Astrophysical Transients with GeV Neutrinos}

\author[0009-0007-2369-258X]{Angelina Sherman}
\affiliation{Department of Physics, Wisconsin IceCube Particle Astrophysics Center, University of Wisconsin, Madison, WI, 53706}

\author[0000-0001-9179-3760]{Jessie Thwaites}
\affiliation{Department of Physics, Wisconsin IceCube Particle Astrophysics Center, University of Wisconsin, Madison, WI, 53706}

\author[0000-0002-5387-8138]{K.~Fang}
\affiliation{Department of Physics, Wisconsin IceCube Particle Astrophysics Center, University of Wisconsin, Madison, WI, 53706}

\author[0000-0002-9867-6548]{Justin Vandenbroucke}
\affiliation{Department of Physics, Wisconsin IceCube Particle Astrophysics Center, University of Wisconsin, Madison, WI, 53706}

\author[0000-0002-4670-7509]{Brian D.~Metzger}
\affiliation{Department of Physics, Columbia University, New York, NY 10027, USA}
\affiliation{Center for Computational Astrophysics, Flatiron Institute, 162 5th Avenue, New York, NY 10010, USA}

\date{\today}

\begin{abstract}
Although Cherenkov detectors of high-energy neutrinos in ice and water are often optimized to detect TeV-PeV neutrinos, they may also be sensitive to transient neutrino sources in the 1-100~GeV energy range. A wide variety of transient sources have been predicted to emit GeV neutrinos. In light of the upcoming IceCube-Upgrade, which will extend the IceCube detector's sensitivity down to a few GeV, as well as improve its angular resolution, we survey a variety of transient source models and compare their predicted neutrino fluences to detector sensitivities, in particular those of IceCube-DeepCore and the IceCube Upgrade. We consider the ranges of neutrino fluence from transients powered by non-relativistic shocks, such as novae, supernovae, fast blue optical transients, and tidal disruption events. We also consider  fast radio bursts and relativistic outflows of high- and low-luminosity gamma-ray bursts. Our study sheds light on the prospects of observing GeV transients with existing and upcoming neutrino facilities. 
\end{abstract}

\section{Introduction}

A wide variety of transient sources are predicted to emit GeV-energy neutrinos. A growing number of optical transients could be powered entirely or in part by non- or mildly-relativistic shocks, which also provide promising sites for proton acceleration and associated hadronic emission \citep{Caprioli_nonrel_shockpowered,2018MNRAS.479..687S, shock_powered_transients, 2022Univ....8...32B}. 
Observational evidence for ion acceleration in shock-powered transients has been accumulated with the recent detections of GeV gamma-ray emission from classical \citep{Ackermann+14} and recurrent novae \citep{Abdo+10,Martin&Dubus13}, the latter also observed to produce TeV emission \citep{MAGIC_hadronic_nova, HESS_hadronic_nova}. A hadronic interpretation for nova gamma-rays is supported by the observed time correlation between the gamma-ray and optical light curves \citep{Aydi_nova_shockpowered}, the short cooling time of putative gamma-ray emitting electrons (disfavoring a leptonic origin; e.g., \citealt{Li_nova_shockpowered}), and the expected ion versus electron acceleration efficiency at non-relativistic shocks \citep{Caprioli_nonrel_shockpowered,Xu+20}. Analogous ion acceleration at shocks may occur under similar physical conditions in a wide array of other extragalactic transients \citep{shock_powered_transients} such as interaction-powered supernovae, tidal disruption events, stellar mergers, and fast blue optical transients \citep{SLSNe_shockpowered, TDE_shockpowered, FBOT_shockpowered}. 

Besides nonrelativistic transients, relativistic outflows of stellar explosions and outbursts are also promising candidate GeV neutrino emitters.  Gamma-ray bursts (GRBs), the most powerful explosive events in the universe, have been proposed as potential cosmic-ray accelerators \citep{2006RPPh...69.2259M}. Moreover, many models postulate that GRBs could produce high-energy neutrinos without relying on cosmic ray acceleration; instead, quasi-thermal neutrinos are produced by internal collisions between differentially streaming protons and neutrons as a result of the decoupling of the electrically neutral neutrons (neutron decoupling scenario; \citealp{Bahcall:2000sa}) or of the  radial differences in the velocities of outflows due to variability in the GRB jet (collision scenario; \citealp{2010MNRAS.407.1033B, Murase:2013hh,  PhysRevLett.110.241101, PhysRevD.105.083023}). 

Among GRBs there exist both high-luminosity (HL) and low-luminosity (LL) varieties, which observations indicate likely comprise two distinct populations of sources (e.g., \citealt{Virgili+09}). LLGRBs may similarly produce GeV neutrinos following the decoupling model of HLGRBs \citep{Murase_LL_GRB, Carpio_LL_GRB, Gupta_HL_LL_GRB}. The GRB population is also divided into long and short ($<2$s) classes. While the collisional and decoupling models may be applicable to short GRBs (neutron star mergers), short GRBs are less well understood and may contain other source engines in addition to jets \citep{Fang:2017tla,Kimura:2017kan,Gottlieb&Globus21}. Therefore, we consider only long GRBs in this work.

Finally, fast radio bursts (FRBs) are millisecond-duration bursts of coherent radio emission \citep{Lorimer+07}. A promising explanation for FRB observations is based on the flaring activity of magnetars, which can produce neutrino emission due to photohadronic interactions of relativistic ions with surrounding synchrotron photons \citep{FRB}. The latter occurs in models where FRB emission is produced by magnetized shocks generated as relativistic ejecta from the magnetar flare collides with some external medium surrounding the FRB site (e.g., \citealt{Metzger+19}).

GeV neutrinos help probe the underlying physics of transient sources as neutrino production is tightly connected to factors such as the composition of the jets, the geometry of the outflow, and the density of their surrounding media. For example, neutrino production in a GRB is highly inefficient unless the jet has a significant baryon content. Furthermore, for neutrino production to occur due to neutron decoupling in a GRB, there must be a substantial neutron to proton ratio in the GRB jet, which in turn depends on the photodisintegration efficiency at early times \citep{GRB_decoupling}, and hence the entropy of the GRB jet \citep{Metzger_GRB_entropy, Ekanger_GRB_nucleosynthesis}. Similarly, the internal shock model for GRB neutrino production depends heavily on the proton acceleration efficiency in the GRB \citep{Murase_LL_GRB}. Finally, neutrino production in the magnetar-flaring model of FRBs depends on the upstream medium being composed of baryons \citep{Metzger+19,FRB}, but would not be present in alternative scenarios where the upstream medium is an electron/positron wind \citep{Beloborodov_FRB121102}. In general the gamma-ray emission from shock-powered transients can arise from either leptonic or hadronic processes, in which case the detection or constraining limits on GeV neutrino emission would help diagnose the dominant emission mechanisms and the physical conditions at the shock.  

The IceCube Neutrino Observatory is a gigaton-scale ice Cherenkov detector located at the geographic South Pole. While the main portion of the detector has sensitivity optimized to the $\sim$TeV-PeV range, the center of the detector, called IceCube-DeepCore, contains a more densely-instrumented region with sensitivity down to $\sim 10$ GeV \citep{GRECO_nova, GRECO_nova_erratum}. The upcoming IceCube Upgrade, which will add further instrumentation and calibration devices inside the IceCube-DeepCore volume, is expected to improve the sensitivity to $\mathcal{O}$(1 GeV) neutrinos \citep{Ishihara:2019uL}. The largest obstacle to GeV neutrino astronomy with IceCube is the neutrino flux generated by cosmic ray interactions in the atmosphere, which overwhelms the signal from most point sources. However, if the time on which a source is observed is sufficiently small, the signal from a transient source could potentially exceed the threshold necessary for detection before a large amount of atmospheric background has time to accumulate. This makes DeepCore and Upgrade promising for astronomy of transients emitting GeV neutrinos.  

IceCube has conducted extensive searches for GeV neutrinos from various transients either by looking for emission by an individual event \citep{GRECO_BOAT} or by stacking the signals from a population of sources \citep{GRECO_GRB, GRECO_nova, GRECO_nova_erratum, GRECO_GW_search, IceCube_transient_search, First_DeepCore_search}. No GeV neutrino transients have been observed to date. Nonetheless, the prospects for GeV transient studies with high energy neutrinos may still be promising. The IceCube-Upgrade and other next-generation facilities will soon provide better sensitivity in the 1-100 GeV range \citep{IceCube_Gen2, ORCA}. In addition, upcoming optical and infrared observatories such as the Vera Rubin Observatory and the Nancy Grace Roman Space Telescope will be able to conduct wider and deeper surveys of the sky, uncovering more transient sources \citep{Rubin_transients, roman_observatory}. These surveys could also be complemented by individual bright events such as a Galactic supernova or nova in the future \citep{next_Galactic_SN}. On the other hand, if further observations at increasing sensitivity continue to yield nondetection of GeV neutrinos from transient sources, this would establish strong constraints on our understanding of the composition and environment of transient sources. Because of this, continued observation of transient sources with improved sensitivity on 1-100 GeV will have fundamental implications for our understanding of many types of astrophysical sources. 

In this paper, we survey models of neutrino emission from a number of transient source classes to estimate their predicted neutrino fluence (time-integrated energy flux) and timescale suitable for observation. We compare these results to the sensitivity of IceCube-DeepCore and projected sensitivity of IceCube Upgrade. In Section 2 we outline the models used for our fluence estimates. In Section 3 we estimate the sensitivities of Upgrade and DeepCore as a function of observation time, and in Section 4 we present our results.

\section{Transient Models}

\subsection{Shock-powered transients}

We base our fluence estimate for shock-powered transients on \cite{shock_powered_transients}, which extends observations of classical novae to be applicable to nonrelativistic shocks in supernovae, novae, TDEs, and FBOTs. 
In this model, each transient is modeled as uniform, spherically-expanding ejecta colliding with a stationary external medium. This collision creates a shock moving forward into the stationary medium and a reverse shock travelling back into the ejecta. The shock observables are the optical luminosity curve $L_\mathrm{opt}$, the peak time $t_{\mathrm{pk}}$, and the mean velocity of the shocked ejecta $v_{\rm ej}$. 

During the initial phases of the shock's outward propagation, the external medium is too opaque for radiation to escape the environment. Instead, thermal emission (e.g., UV or X-ray photons) from the shocked gas are absorbed by the ejecta and efficiently reprocessed to optical light. When the optical depth of the material ahead of the expanding shock has dropped below the critical value $\tau_{\mathrm{opt}} \approx c/v_{\mathrm{sh}}$, this reprocessed optical radiation can escape the shock to a distant observer. This transition occurs at the critical time $t_{\mathrm{pk}}$, which coincides with the peak of the transient's optical light curve. Because of this, after time $t_{\mathrm{pk}}$ the kinetic power of the shock should faithfully track the observed optical luminosity: $L_{\mathrm{sh}} \approx L_{\mathrm{opt}}$. We thus make the optimistic assumption that all of the transient's light is powered by reprocessed shock emission, with only small contributions from other sources of luminosity potentially present in these sources, such as radioactivity or a central compact object. 

Not only is the optical emission which escapes from the shock suppressed at times earlier than $t_{\mathrm{pk}}$, but relativistic ion acceleration is also difficult because the shock transition is mediated by radiation instead of being collisionless during these phases (see \citealt{Levinson&Nakar20} for a review of radiation-mediated shocks). Instead, the bulk of the particle acceleration must occur after $t_{\mathrm{pk}}$. The accelerated ions during this latter phase carry power $L_{\mathrm{rel}} \approx \epsilon_{\mathrm{rel}} L_{\mathrm{sh}}$, where $\epsilon_{\mathrm{rel}}$ is the is the particle acceleration efficiency. In classical novae, this efficiency is measured to be $\epsilon_{\mathrm{rel}} \sim 0.003 - 0.01$  \citep{Metzger_gammaray_nova, Aydi_nova_shockpowered}, based on the observed ratio of gamma-ray to optical luminosities (the gamma-ray luminosity tracks the fraction of the shock power into non-thermal particles while the optical luminosity tracks the total shock power); in our fluence estimates we consider a range of efficiencies $\epsilon_{\mathrm{rel}} = 0.01 - 0.1$, which bracket the measured values in novae and the $\sim 10\%$ maximum ion acceleration efficiency predicted for non-relativistic shocks by particle-in-cell simulations \citep{Caprioli_nonrel_shockpowered}.

We assume that protons accelerated at the shock follows a standard power-law energy spectrum $\frac{dN}{dE_p} \propto E_{\text{p}}^{-\alpha}$
where $\alpha$ is the spectral index (e.g., \citealt{2018MNRAS.479..687S}). We consider a range of values $\alpha = 2-2.7$. The proton spectrum is normalized by the total energy output over the transient's duration $\Delta t$, viz.~
\begin{equation}
    \int dE_p E_p \frac{dN}{dE_p} =\int_{t_{\text{pk}}}^{t_{\rm pk} + \Delta t} L_{\rm rel} \text{dt}. 
\end{equation}
We estimate $\Delta t$ as the time from the peak luminosity $t_{\rm pk}$ to when when the shock power/neutrino luminosity has dropped to half its peak value. For the transient classes we consider, this timescale ranges from a few days to novae to a few months for TDEs.
 
After leaving the vicinity of the shock, the accelerated protons collide with other ambient ions, generating pions which subsequently decay into neutrinos and gamma-rays. The neutrino $E^2 dN/dEdA$ from a nearby (i.e., non-cosmological) source at distance $d_{\rm min}$ can be estimated by: 
\begin{equation}
\label{shockpowered_fluence}
E_{\nu}^2 \frac{dN}{dE_\nu dA} =  \frac{1}{2}E_{p}^2 \frac{dN}{dE_p} f_{\rm pp} \frac{1}{4 \pi d_{\mathrm{min}}^2},
\end{equation}
where the factor of 1/2 arises because charged pions are produced with 2/3 probability in proton-proton interactions, and about 3/4 of the charged pion energy is carried away by neutrinos. Here, $f_{\rm pp} = 1 - \exp (-t_{\rm dyn}/t_{\rm{pp}}) = 1 - \exp (-\tau_{\rm pp}c/v_{\rm{sh}})$ is the pion production efficiency, $\tau_{\mathrm{pp}}\approx n_{\mathrm{sh}}\sigma_{\mathrm{pp}}R_{\mathrm{sh}}$ is the optical depth to proton-proton production, 
where $\sigma_{\mathrm{pp}}$ is the inelastic proton-proton cross-section. Following the discussion above, we normalize $n_{\text{sh}}$ such that the peak emission timescale $t_{\rm pk}$ equals the critical ``Arnett'' timescale, at which the photon diffusion time through the ejecta external to the shock and the adiabatic loss time are equal (\citealt{Arnett82}). After the peak $t > t_{\rm pk}$, the radial profile of $n_{\text{sh}}$ encountered by the shock is chosen such that the kinetic luminosity of the shock matches the transient's optical light curve, under the assumption of a constant radial shock speed.

We evaluate the spectra of secondary neutrinos using \texttt{Aafragpy} \citep{Aafragpy}. In particular, we use the model from \citet{AAfrag} to evaluate the $pp$ cross-section above 4 GeV, and we use the model from \citet{Kamae_2006} for the cross-section below 4 GeV.

We divide the shock-powered transients into five broad source classes: novae, luminous red novae (LRNe), supernovae (SNe), tidal disruption events (TDEs), and fast blue optical transients (FBOTs). For each case, we employ a light curve ``template" normalized to the peak luminosity $L_{\rm pk}$ and timescale $t_{\rm pk}$ of each transient subtype. The adopted ranges of volumetric rates, peak luminosity, and ejecta velocity for each follow those compiled in Table 1 of \citet{shock_powered_transients}, while $t_{\rm peak}$ is taken as the mean of the given range.

\textit{Novae}: A number of novae have been detected over the past $\sim$15 years in GeV-TeV gamma-rays. These include over a dozen classical novae \citep{Li_nova_shockpowered, Aydi_nova_shockpowered, LAT_novae_2014, LAT_novae_2016} as well as a few recurrent novae \citep{Martin&Dubus13}, most recently RS Ophiuchi  \citep{MAGIC_hadronic_nova, HESS_hadronic_nova, LAT_Ophiuci}.  For classical novae we use as a light curve template of a nova eruption on a solar-mass white dwarf from \citet{nova_light_curve}. We use the volumetric rates from \citet{Galactic_nova_rate} and  \citet{Kochanek_2014}, considering separately events in our Galaxy versus extragalactic novae.  For Galactic rates, we approximate the Galaxy as a thin disk of radius 20 kpc centered about the Sun. Based on the Galactic nova rate of 27 - 81 $\text{yr}^{-1}$ from \citet{Galactic_nova_rate}, we infer an event surface density $\Sigma_{0, \text{GN}}$ from which we estimate the distance to the nearest Galactic event $d_{\text{min}} = \big(\pi t_{\text{obs}} \Sigma_{0, \text{GN}} \big)^{-1/2} \text{Gpc}$, over a given observing duration $t_{\rm obs}$.

We note that the TeV gamma-ray emission in recurrent novae such as RS Oph 
\citep{MAGIC_hadronic_nova, HESS_hadronic_nova, LAT_Ophiuci} may originate from distinct shocks, located at larger distances from the white dwarf, than the more compact shocks which produce the GeV emission \citep{Diesing_RS_manyshocks}. Because of the lower density of the shocked gas in this case, the optical light curve may no longer directly trace the shock power contributing to the TeV emission. For this reason, and because the focus of this work is on GeV emission, we omit this emission component from our modeling.

\textit{Luminous red novae (LRNe)}: Similar to classical novae, Luminous red novae (LRNe) from stellar mergers are likely at least in part powered by shock interaction \citep{Metzger&Pejcha17,Matsumoto&Metzger22}. For this source class, we adopt the light curve of V1309~Sco as the template \citep{2010A&A...516A.108M, 2011A&A...528A.114T, Matsumoto&Metzger22} and a volumetric rate of  0.2 $\text{yr}^{-1}$ per Galaxy from \citet{Galactic_LRNe_rate}. V1309~Sco was a Galactic LRN with a firm association with merging binary stars. It is among the dimmest but volumetrically most common type of LRNe.   

\textit{Supernovae (SNe)}: To date, nonthermal gamma-ray emission has not been observed from supernovae (SNe) \citep{LAT_SN_nogamma15, LAT_SN_nogamma18}. It has been suggested that a dense circumstellar medium around SNe could attenuate the TeV gamma-ray signal, which would conceal the signature of internal shocks \citep{Cristofari_SN_gamma, Andrews_SN_gamma}. Because of this, SNe are still anticipated to be significant sources of high-energy neutrinos \citep{Murase_SN_predictions}.  We consider all core-collapse supernovae (CCSNe) as a subset, and we additionally consider the class of SNe IIn only, which are a subset of CCSNe with clear evidence for shock interaction. The CCSNe rate is obtained from \citet{CCSNe_rate}, and the rate of SNe IIn is taken to be 8.8\% of the CCSNe following \citet{SNeIIn_rate}. We also evaluate the fluence of both Type I and Type II superluminous supernovae (SLSNe), which are SNe with very high optical luminosity \citep{SLSNe_shockpowered}. The SLSNe rates are obtained from \citet{SLSNe_rates}. Finally, a small subset of Type Ia SNe (Type Ia CSM) show evidence for shock powering \citep{Bochenek_IaCSM}, and we estimate their fluence assuming they represent 0.1-1\% of the SNe Ia rate \citep{IaCSM_rate}. For our supernova light curve template, we use the light curve of a typical SLSNeII from \citet{Inserra_2019}.

\textit{Tidal Disruption Events (TDEs)}: Tidal disruption events (TDEs) are optical and UV transients \citep{Stone_2015_TDE, Arcavi_TDE_obs, Gezari_TDE_obs} whose light curves may be powered by shocks \citep{piran_TDE_shocks, Jiang_TDE_shocks}. TDEs are predicted to be emitters of high energy neutrinos and gamma-rays \citep{Dai:2016gtz,Guepin:2017abw,Winter:2022fpf,Murase_TDE}. We use the light curve template of the TDE event PTF09ge from \citet{TDE_light_curve}. The rate of TDEs is consistent with \citet{TDE_rate} and \citet{vanVelzen_TDE}; the range of peak luminosities is obtained from \citet{vanVelzen_TDE}, and the peak time is obtained from \citet{TDE_tpk}. 

\textit{Fast Blue Optical Transients (FBOTs)}: Fast blue optical transients (FBOTs) are fast, luminous, UV-bright transients that exhibit characteristics similar to SNe, but have timescales and luminosities inconsistent with traditional SN models \citep{FBOT_observations, FBOT_CSS16}. The class of sources producing FBOTs is unknown, and some models postulate shock powering as an explanation for their observed characteristics \citep{FBOT_shockpowered}. High-energy neutrinos from FBOTs have been studied in \citet{FBOT_light_curve}. The nearby event AT2018cow provides a particularly well-studied case of such a transient \citep{AT2018cow} and we use its light curve as our template. The rates and peak time of FBOTs are derived from \citep{FBOT_observations}. We consider separately the category of especially luminous FBOTs ($M_g < -19$), of which AT2018cow is an example. For luminous FBOTs, the rate is taken from \citet{FBOT_CSS16} and the peak time is based on \citet{FBOT_tpk}.

\subsection{Gamma Ray Bursts}

In this section, we estimate the expected fluence of high- (HL) and low-luminosity (LL) GRBs. We consider both decoupling  and collisional scenarios for the HLGRB, and the decoupling scenario for the LLGRB. We obtain the total expected fluence for each case following \citet{Murase_BOAT}, which applies the decoupling and collisional models to the recent event GRB 221009A; we reiterate their fluence calculation below. The parameters that we input to our GRB models are the bulk Lorentz factor $\Gamma$ and isotropic equivalent gamma-ray energy $\mathcal{E}_{\gamma, \text{iso}}$, as well as GRB rates. 

In the neutron decoupling scenario, the neutron and proton populations in the GRB jet are initially coupled by nuclear elastic scattering. As the bulk $np$ outflow expands, the scattering time eventually exceeds the comoving expansion time, at which point the neutron population decouples from the proton population. If the decoupling occurs while the outflow is still accelerating (before $\Gamma = \Gamma_{\mathrm{max}}$), the proton flow continues to be accelerated by proton coupling to photons, while the neutron flow begins to coast. In this way, the neutron and proton flows can acquire a drift velocity, causing inelastic $np$ collisions. 

The energy of neutrinos produced in the decoupling process is given by:
\begin{equation}
\label{decoupling_E}
E_{\nu} \approx 0.1 \Gamma_{n, \mathrm{dec}} m_pc^2/(1+z).
\end{equation} 
Here, $\Gamma_{n, \mathrm{dec}}$ is the Lorentz factor of neutrons at decoupling, and it is estimated to be  
\begin{equation}
    \Gamma_{n, \rm dec} = \frac{3}{4}\left(\frac{L_p\sigma_{np}\Gamma_*}{4\pi R_* \Gamma_{\rm max} m_p c^3 }\right)^{1/3},
\end{equation}
where $\Gamma_* = 10$ and $R_* = 10^{11}\,\rm cm$ are breakout Lorentz factor and injection radius, which have been set to fiducial values, and $\sigma_{np}\approx 3\times 10^{-26}\,\rm cm^2$ is the neutron-proton cross section. The proton luminosity $L_p$ is evaluated as $L_p \sim \xi_N {\cal{E}}_\gamma^{\rm iso} / \Delta t$. We use a range of nucleon loading factors $\xi_{N} = 3-30$ as suggested by \citet{Murase_BOAT}, and we account for this range of loading factors in the uncertainty of our final fluence calculation. 
The total flavor-summed $(\nu + \overline{\nu})$ neutrino fluence is estimated as:
\begin{equation}
    \label{decoupling_fluence}
   \int dE_\nu\, E_{\nu} \frac{dN}{dE_\nu dA} \approx \frac{1}{4} \frac{(1+z)}{4 \pi d_{L}^2} \zeta_n \bigg(\frac{\Gamma_{n, \mathrm{dec}}}{\Gamma}\bigg) \xi_{N}\mathcal{E}_{\gamma}^{\mathrm{iso}}.
\end{equation}
Where we assume that the number ratio of protons to neutrons is $\zeta_{n} = 1$. In $np$ collisions, 2/3 of the produced pions are charged, and 3/4 of their decay products are shared by each flavor of neutrinos. Additionally, the nucleon inelasticity of $np$ collisions is $\approx 0.5$, altogether contributing the factor of 1/4. 

Because the GRB jet itself is variable, neutrons from a slower region may also diffuse into a faster region of the jet, creating a compound flow with $\Gamma_n < \Gamma_p$, resulting in neutrino production from inelastic $np$ collisions. These collisions can occur even in cases where the $np$ decoupling occurs after $\Gamma = \Gamma_{\mathrm{max}}$ is achieved, and the relative velocities between the proton and neutron flows are not sufficient to pion-produce in $np$ collisions from decoupling alone. 

The expected energy of neutrinos produced in the collisional process is: 
\begin{equation}
\label{collision_E}
E_{\nu} \approx 0.1\Gamma\Gamma_{\mathrm{rel}}'mpc^2/(1+z)
\end{equation}
giving 30-300 GeV neutrinos. Here, $\Gamma_{\mathrm{rel}}' \sim 2$ is the relative Lorentz factor of the interacting flow. The flavor-summed $(\nu + \overline{\nu})$ neutrino fluence is given by:
\begin{equation}
\label{collision_fluence}
\int dE_\nu\, E_{\nu} \frac{dN}{dE_\nu dA } \approx \frac{1}{4} \frac{(1+z)}{4 \pi d_{L}^2} \xi_{N}\tau_{pn}\mathcal{E}_{\gamma}^{\mathrm{iso}}
\end{equation}
where we assume the $pn$ optical depth is $\tau_{pn} = 1$. 

As in the case of shock-powered transients, we evaluate the spectra of GRB neutrinos using \texttt{Aafragpy} \citep{AAfrag, Aafragpy, Kamae_2006}. For the decoupling scenario, we evaluate the neutrino spectra from a monoenergetic spectrum of rest-mass energy protons. In the collisional scenario, where a wider range of proton energies can contribute to neutrino production, we use a primary proton spectrum following a Maxwell-Boltzmann distribution with a mean energy of the proton rest mass: 
\begin{equation}
\frac{dN}{dE_\text{p}} \propto 2\sqrt{\frac{E_p}{\pi}}\bigg(\frac{1}{\text{kT}}\bigg)^{3/2}e^{-E_{\text{p}}/\text{kT}}
\end{equation}
with $\text{kT} = \frac{2}{3}m_{\text{p}}c^2$.
 
The spectra are normalized such that the total fluence 
is equivalent to the values given by Equations \ref{decoupling_fluence} and \ref{collision_fluence}. We normalize the spectral peaks so that the peak energy aligns with the values predicted by Equations \ref{decoupling_E} and \ref{collision_E}. We outline the parameters used for high- and low-luminosity GRBs below.

\textit{High-luminosity GRBs}: For HLGRBs, we consider neutrino production both in the decoupling and collisional scenarios. We consider ranges $\Gamma = 100 - 1000$ and $E_{\gamma, \text{iso}} = 10^{52} - 10^{54}$ erg \citep{Atteia_Egamma_iso, Murase_BOAT, GRB_decoupling}, as well as an observational time window $\Delta t = 100$s, based on T90 values presented in \citep{Atteia_Egamma_iso}. We use a range of true GRB rates $R_{0, \text{HL}} = 0.5 - 2 \text{ Gpc}^{-3}\text{yr}^{-1}$ \citep{Swift_GRB_rate}, which gives a conservative estimate of the nearest high-luminosity GRB which may produce neutrinos detectable at Earth. 

\textit{Low-luminosity GRBs}: LLGRBs likely comprise a distinct population of sources from their HL-counterparts. However, the decoupling model from HLGRBs is often considered when modeling LLGRBs. For LLGRBs, we use $\Gamma = 10 - 100$ and $E_{\gamma, \text{iso}} = 10^{49} - 10^{51}$ erg \citep{Carpio_LL_GRB, Murase_LL_GRB}, and observational time window $\Delta t = 20$s \citep{Nakar_LL_GRB_duration, Piran_LL_GRB_duration}.We use a range of geometrically-corrected LL
GRB rate of $R_{0, \text{LL}} = 200 - 500 \text{ Gpc}^{-3}\text{yr}^{-1}$ following \citet{Murase_LL_GRB}.

\begin{figure*}[t]
    \centering
    \includegraphics[width =0.9 \linewidth]{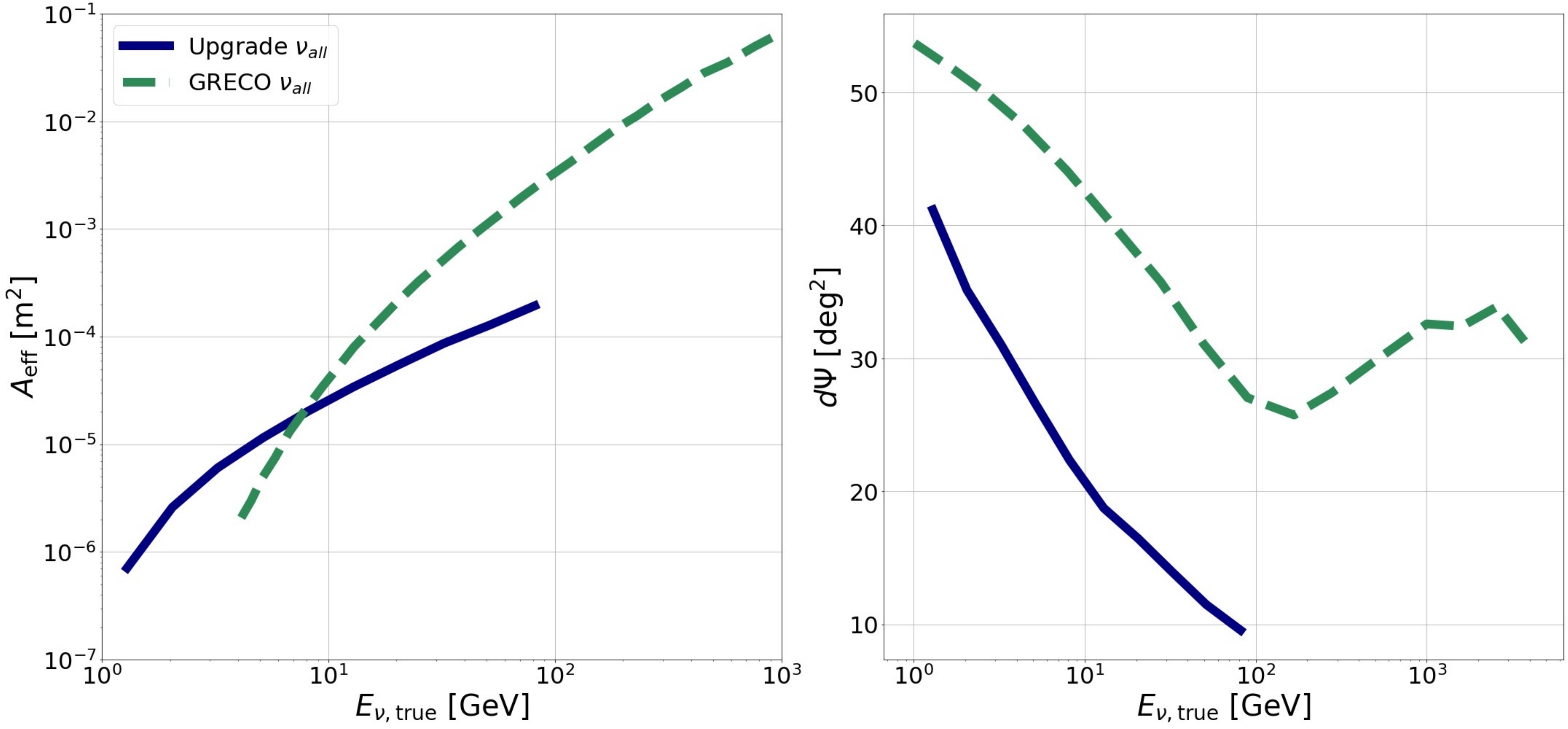}
    \caption{\label{detector_specs} Flavor-summed $(\nu + \overline{\nu})$ effective area and angular resolution of GRECO and Upgrade. The GRECO dataset is optimized on 10 GeV - 1 TeV; Upgrade is intended to provide additional sensitivity to events on 1 - 10 GeV. We obtain GRECO Northern sky effective area and median angular resolution from \citet{GRECO_nova_erratum}. We evaluate the expected Upgrade parameters using simulated data provided by \citet{Icecube_Upgrade_data}. The effective area for GRECO is averaged over the northern sky, while the effective area for Upgrade is averaged over the full sky.} 
\end{figure*}

\subsection{Fast Radio Bursts}

We base our FRB fluence estimates on \cite{FRB}, who calculate the neutrino emission from a shock generated by ejecta from a magnetar flare colliding with external baryon-loaded medium (e.g., an earlier ejecta shell from the same magnetar; \citealt{Metzger+19}). The model is motivated in part by observations of the Galactic FRB 200428, which was observed in coincidence with an X-ray outburst from the Galactic magnetar SGR 1935+2154 \citep{FRB200428}. To date, this is the only FRB observed from our Galaxy with properties broadly consistent with those of extragalactic FRBs (typically at cosmological distances), and its fluence was still on the very low end of this range. Nevertheless, we focus in this paper on the detection prospects for a Galactic FRB but with properties more energetic and hence typical of the cosmological FRB population than FRB 200428.

We consider only neutrino production from the thermal population of ions heated at the relativistic shock (as opposed to neutrinos from non-thermal ions accelerated by the shock, which generate higher-energy $\gtrsim$ TeV neutrinos; \citealt{FRB}). In the following discussion, the primed frame refers to the comoving frame of the relativistic outflow. The FRB observables are its duration $t_{\mathrm{FRB}}$, radio frequency $\nu_{\mathrm{obs}}$, and radio energy $\mathcal{E}_{\mathrm{radio}}$. These can be related to the shock radius $r_{\mathrm{FRB}}$, density $n_{\mathrm{FRB}}$, Lorentz factor of the shock $\Gamma_{\mathrm{FRB}}$ and luminosity of the FRB $L_{\rm FRB}$ through equations~5-10 of \citet{FRB}. 

In the case of a uniform density profile of the external medium, the shock parameters evolve with time as: 
\begin{equation*}
   \Gamma = \Gamma_{\mathrm{FRB}} \bigg(\frac{t}{t_{\mathrm{FRB}}}\bigg)^{-3/8} 
\end{equation*}
\begin{equation*}
r_{\mathrm{sh}} = r_{\mathrm{FRB}} \bigg(\frac{t}{t_{\mathrm{FRB}}}\bigg)^{1/4}
\end{equation*}
\begin{equation*}
L_{\mathrm{sh}} = L_{\mathrm{FRB}} \bigg(\frac{t}{t_{\mathrm{FRB}}}\bigg)^{-1}.
\end{equation*}

We assume that the shock's energy is shared equally between protons and electrons in the plasma, where both are heated to Maxwellian energy distributions \citep{Sironi_plasma}. The heated electrons emit synchrotron radiation, which peaks at photon energy: 
\begin{equation}
    \epsilon_{\rm{pk}} = \frac{\hbar e B'}{m_e c}\overline{\gamma}^2 \Gamma
\end{equation}
which is on the order of hundreds of MeV. Here, $B' = \sqrt{64\pi\sigma\Gamma^2 m_p c^2 n_{\rm{FRB}}}$ is the post-shock magnetic field and $\overline{\gamma} = (m_p/m_e)\Gamma/2$ is the mean thermal Lorentz factor.

We consider neutrino production from a distribution of thermal protons with mean energy $\overline{E}_p' = \Gamma m_pc^2/2$ immersed in a field of monoenergetic photons at energy $\epsilon_{\gamma}' = \epsilon_{\rm{pk}}/\Gamma$. 
The energy of protons that interact with the synchrotron photons is evaluated as $E'_p= E'_\Delta m_p c^2 / (2 \epsilon'_\gamma )$, where $E'_\Delta \approx 0.3$~GeV is the $\Delta$-resonance energy. The pion production by thermal protons lasts until $\epsilon'_\gamma$ drops so much that the threshold condition is no longer satisfied for mean-energy thermal protons. Beyond that time, pion production can only occur to non-thermal protons. We use this time $t_{\mathrm{nth}}$ as the value for $\Delta t$ over which the FRB is observed. This timescale, which is much shorter than for any other transient source, falls into a fully background-free regime for IceCube observation.

The proton energy spectrum follows a relativistic Maxwell-Boltzmann distribution:
\begin{equation}
    \frac{dN}{dE_p} \propto \frac{dN}{d\gamma_p} = \frac{\big(\frac{\gamma_p}{\Gamma}\big)^2 e^{-\gamma_p/\Gamma\Theta}}{2 \Gamma \Theta^3},
\end{equation}
where $\gamma_p = E_p/m_pc^2$ and $\Theta = \frac{1}{3}(\Gamma/2)$. The distribution is normalized to reflect the total shock energy conferred to protons during the thermal period, or  $\frac{1}{2}\int^{t_{\mathrm{nth}}} L_{\mathrm{sh}} dt$. The neutrino $E_\nu^2 dN/dE_\nu dA$ from a FRB at distance $d$ is estimated by
\begin{equation}
\label{FRB_fluence}
    E_{\nu}^2 \frac{dN}{dE_\nu dA} = \frac{1}{4\pi d^2} \frac{3}{8} \tau_{p\gamma} E_{p}^2 \frac{dN}{dE_p},
\end{equation}
where the factor of $3/8$ accounts for the fraction of proton energy that goes into neutrinos during a $p\gamma$ interaction. The optical depth of $p\gamma$ interaction is given by $\tau_{p\gamma}\approx n'_\gamma \sigma_{p\gamma}r'$, where $\sigma_{p\gamma}$ is the inelastic $p\gamma$ interaction cross section, $n'_\gamma\approx u'_\gamma / \epsilon'_{\rm{pk}}$ and $r' \approx r_{\rm sh}/\Gamma$ are the photon number density and radius of the post-shock region in the co-moving frame, respectively. The radiated power is estimated as $u'_\gamma \approx (\frac{1}{2} L_{\rm sh}) / (4\pi r_{\rm sh}^2 c \Gamma^2)$ considering that about half of the shock power goes to thermal electrons. Note that in a typical FRB, the thermal proton energy is too high for Bethe-Heitler pair production to be significant. The energy of a synchrotron photon in the rest frame of a thermal proton is on the order of hundreds of MeV, where the effective cross section (product of cross section and inelasticity) is very suppressed.

For our estimate, we use $\mathcal{E}_{\text{radio}} = 10^{37} - 10^{41}$ erg and $t_{\text{FRB}} = 0.1 - 10$ ms, which are typical for most extragalactic FRBs. We use an optimistic distance $d = 10$ kpc, which is the distance of Galactic FRB 200428. FRB 200428 was the only observed Galactic FRB with parameters comparable to the extragalactic FRB population, and the Galactic FRB rate is poorly constrained. The event FRB 200428 had a radio energy of $10^{35}$erg, so our estimate is quite optimistic and depends on a more typical FRB occuring in our Galaxy during the next decade.

\section{Estimation of IceCube sensitivity}

In this section, we estimate the sensitivities of IceCube Upgrade and DeepCore as a function of observation time. Our approach to estimating sensitivity is an analytic approximation that is intended to give perspective on the potential detection prospects for various transient types, and should not be considered to be a formal evaluation of IceCube's sensitivity. In particular, a formal estimation of IceCube's sensitivity would use a log-likelihood method over the full energy range of the dataset. In addition, such an estimate would consider event-by-event angular resolution rather than the median angular resolution, and would evaluate the sensitivity for a source at a specific declination, while our estimate considers an average sensitivity over a broad range (averaged over the Northern hemisphere for GRECO and all-sky averaged for Upgrade). Such targeted searches may improve the sensitivity over the estimates shown here. Nevertheless, our analytic estimate is a reasonable proxy of the IceCube sensitivity for the broad, orders-of-magnitude comparisons we make in this analysis. Our method for estimating the sensitivities of IceCube-DeepCore and the IceCube Upgrade are outlined below. 

We obtain atmospheric neutrino flux estimates from \citet{Honda_atmosphere}. We evaluate the total, flavor-summed  number of observed atmospheric neutrinos using: 
$$N_{\text{atm}} = t_{\text{obs}}\int_{E_{\text{min}}}^{E_{\text{max}}}\frac{dN}{dEdAd\Omega dt} A_{\text{eff}}\Omega dE,$$
where $E_{\mathrm{min}}, E_{\mathrm{max}}$ give the energy range of observation, $A_{\mathrm{eff}}$ is the effective area of the detector, $t_{\mathrm{obs}}$ is the observation time, and $\Omega = d\Psi^2$ is the solid angle of the detector's resolution, which is evaluated using a median angular resolution $d\Psi$. For sufficiently low observation times, there may be no observed background events; the time interval on which this condition is met is considered the ``background-free" regime. 

Our estimate of the sensitivity follows the formalism for the case of a Poisson distribution with known background described in \citet{Kashyap_2010}, and which we briefly outline here. For observation over a given time interval $t_{\text{obs}}$, we constrain the probability of a false positive detection $\alpha$ using:
$$\text{Pr}(N_s > N_s^*|\lambda_s = 0, \lambda_{\text{atm}}, t_\text{obs}) = \frac{\gamma(N_s^* + 1, N_{\text{atm}})}{\Gamma(N_s^* + 1)} \leq \alpha$$
where $\alpha = 0.1$, $N_{\text{atm}} = \lambda_{\text{atm}}t_{\text{obs}}$, and $N_s^*$ gives the number of events at the threshold. Here $\gamma(a, x)$ is the incomplete gamma function as defined in equation 8.350.1 of \citet{Gamma_functions}. We constrain the probability of detection at the threshold $\beta$ using:
$$\text{Pr}(N_s > N_s^*|\lambda_s, \lambda_{\text{atm}}, t_\text{obs}) = \frac{\gamma(N_s^* + 1, N_s^* + N_{\text{atm}})}{\Gamma(N_s^* + 1)} = \beta$$
taking $\beta = 0.9$. This yields the number of signal events $N_{s, 90}$ necessary for a source to exceed the background-only expectation over a given observation time $t_{\text{obs}}$. 

We assume a source power-law spectrum $dN/dE \propto E^{-s}$, where s is the source spectral index. This gives an expected number of observed signal neutrinos:
$$N_s = K \int_{E_{\text{min}}}^{E_{\text{max}}} E^{-s} A_{
\text{eff}}dE,$$
where the coefficient K is a normalization factor. Taking $K = N_{s, 90}/N_s$ gives the spectrum at the detection threshold. The minimum fluence (time-integrated energy flux) necessary for a neutrino source to be detected over the atmospheric background is then: $$F_{\nu, \mathrm{lim}} = K\int_{E_{\mathrm{min}}}^{E_{\mathrm{max}}}E^{(1-s)}dE.$$

For both GRECO and Upgrade, we consider the effective area for all neutrino flavors $(\nu + \overline{\nu})$. To estimate the sensitivity of IceCube-DeepCore, we use the GRECO (GeV Reconstructed Events with Containment for Oscillation) dataset. This dataset is optimized for observations on 10 GeV - 1 TeV, and is often used to search for low-energy transients. In the case of GRECO, we use an effective area averaged over the northern sky (which gives a slightly more optimistic sensitivity estimate than the southern sky). We obtain the GRECO effective area and angular spread of events $\Delta\Psi$ from \citet{GRECO_nova_erratum}; we multiply the effective area presented in \citet{GRECO_nova_erratum} by 2 to give an flavor-summed $(\nu + \overline{\nu})$ estimate, and we use the median value of $\Delta\Psi$. 

The upcoming IceCube Upgrade, which will add further instrumentation and calibration devices to IceCube DeepCore, is expected to improve the sensitivity to $\mathcal{O}$(1 GeV) neutrinos. We obtain a projection of the effective area and angular resolution of Upgrade using simulated IceCube-Upgrade detector data provided by \citet{Icecube_Upgrade_data}. For the case of IceCube-Upgrade, we consider effective area averaged over the whole sky. It should be noted that the estimates of effective area and angular resolution presented in \citet{Icecube_Upgrade_data} should be considered preliminary, and these values may be further optimized in the future. We show the effective area and angular resolutions of GRECO and Upgrade in Figure \ref{detector_specs}. We evaluate the sensitivities assuming source spectral indices ranging from $s = 1.5$ to $s = 3$.

\section{Results}

The ranges of the parameters used in our fluence predictions for various transient classes are summarized in Table \ref{transient_values}. For extragalactic events, we estimate the nearest event within $t_{\text{obs}}$ years of IceCube observation using the event's the volumetric birth rate $R_{0}$ $\text{Gpc}^{-3}\text{yr}^{-1}$ as:
\begin{equation}
   d_{\text{min}} \sim \left(\frac{4\pi}{3}t_{\text{obs}} R_{0} \right)^{-1/3} \text{Gpc}.
\end{equation}

\begin{figure}[h]
    \centering
    \includegraphics[width = 1\linewidth]{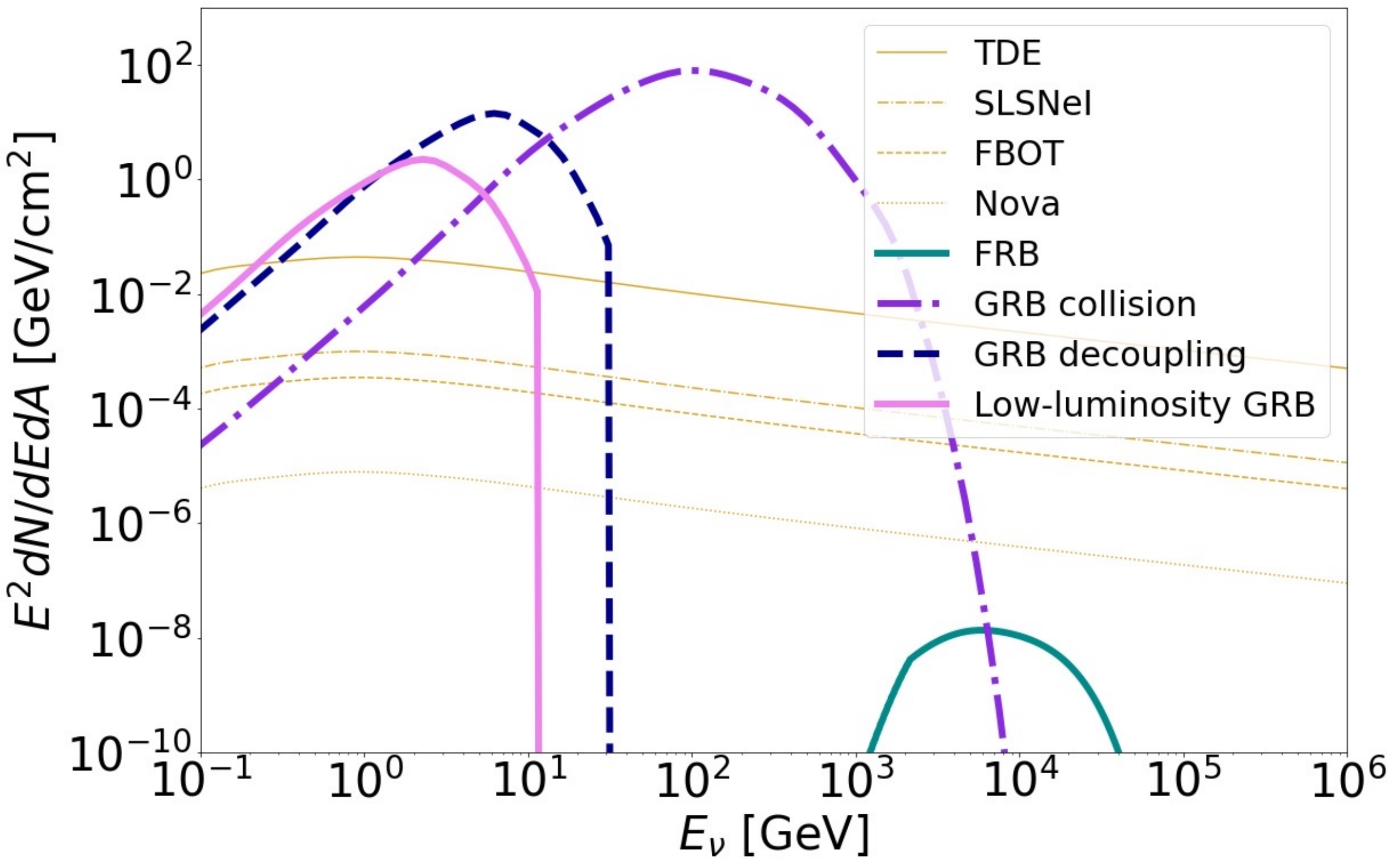}
    \caption{\label{spectra} $E^2 dN/dEdA$ of various transient sources. The spectra of four shock-powered transient source types are shown with spectral index $\alpha = 2.4$, and are normalized to the mean values shown in Table \ref{transient_values}. The FRB spectrum for thermal neutrino emission is evaluated using fiducial parameters $t_{\mathrm{FRB}} = 1$ ms and $\mathcal{E}_{\mathrm{radio}} = 10^{40}$ erg, and $d = 10$ kpc. In the case of a high-luminosity GRB, the spectra are shown separately for the decoupling and collision models; in the case of a low-luminosity GRB, the spectrum is shown for the decoupling model only. All GRB spectra are normalized using mean values from Table \ref{transient_values}.}     
\end{figure}

Figure \ref{spectra} shows time-integrated, energy-scaled spectra $(E^2 dN/dEdA)$ of several transients considered in this work. The spectra shown use mean values from Table \ref{transient_values}. For shock powered transients, the spectral index $\alpha = 2.4$ is shown, although in our fluence estimate we consider the range of values $\alpha = 2 -2.7$. The expected fluence of a transient on a particular energy interval is given by integrating its (already time-integrated) energy spectrum over the desired energy interval: 
\begin{equation}
\label{fluence_integral}
F_{\nu} =  \int_{E_{\text{min}}}^{E_{\text{max}}}dE_\nu\, E_\nu \frac{dN}{dE_\nu dA}.
\end{equation}

For perspective, we also show an optimistic estimate of a Galactic core-collapse supernova at 10 kpc. The true rate of Galactic CCSNe is $\sim$ 3 per century \citep{next_Galactic_SN}, so there is only a  $\sim$ 30\% chance of such an event occuring in the Galaxy in the next decade. However, if such an event were to occur, the prospects for detection could be favorable.

\begin{figure*}[t]
    \centering
    \includegraphics[width = 0.9\linewidth]{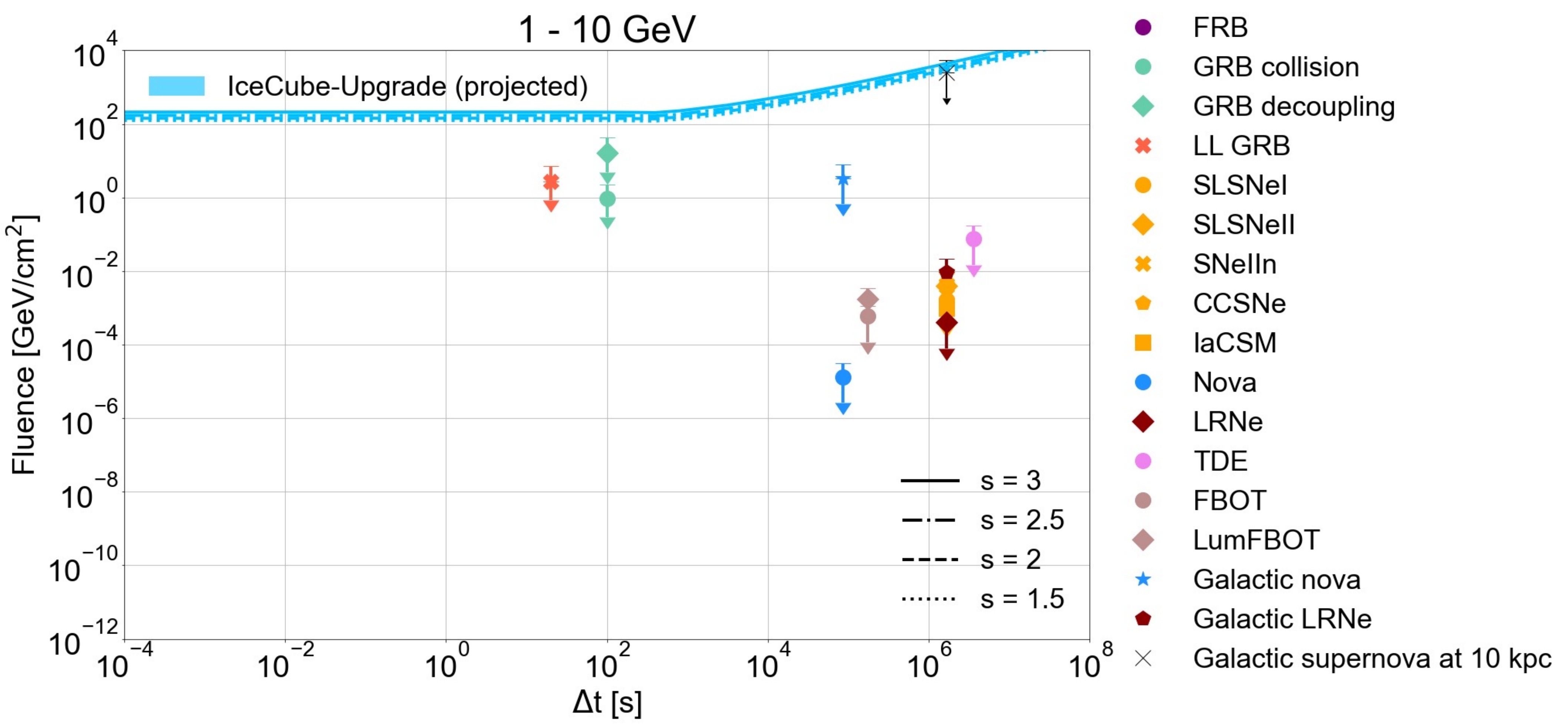}
    \includegraphics[width = 0.9\linewidth]{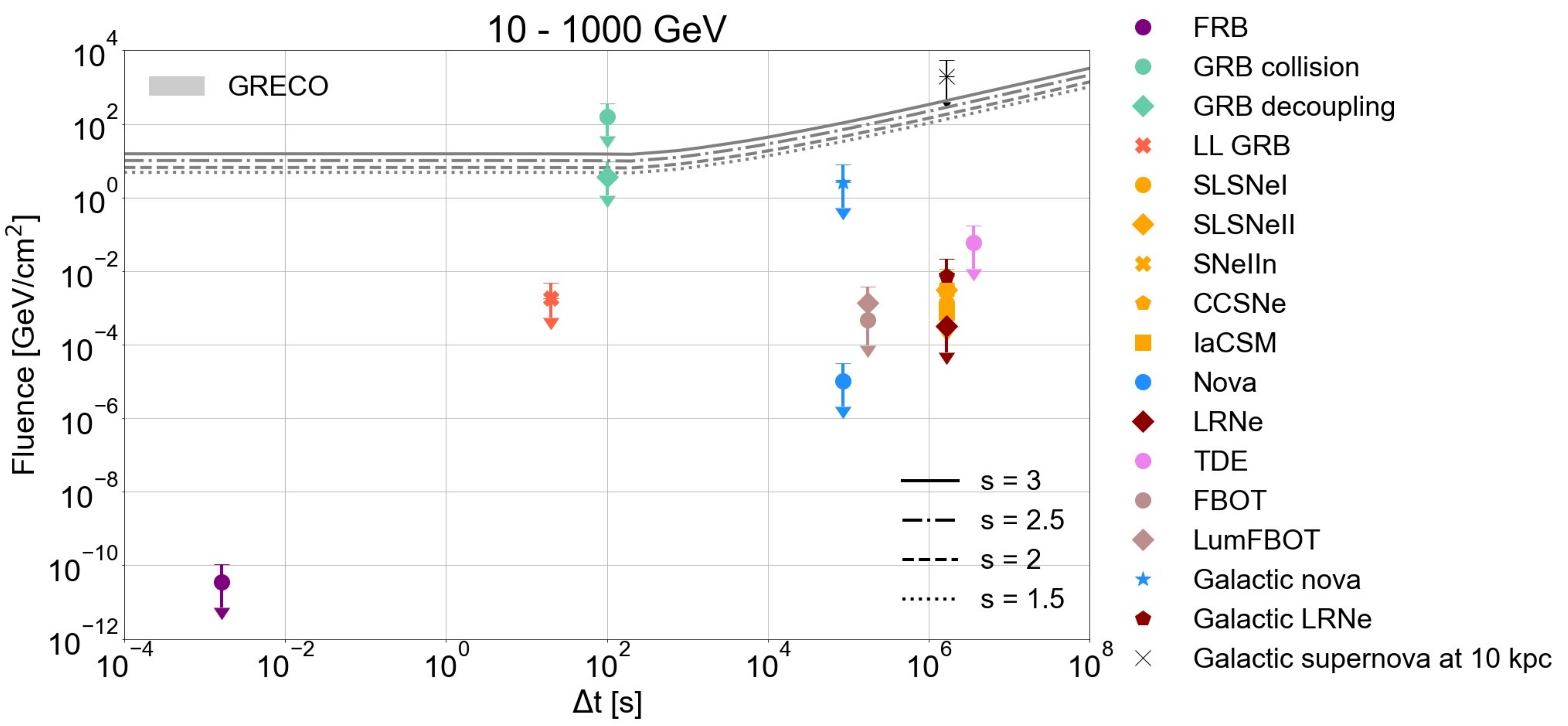}
    \caption{\label{fluence_estimate} Predicted source fluences (time-integrated energy flux) and IceCube sensitivities integrated over 1-10 GeV (above) and 10 - 1000 GeV (below). Downward arrows indicate error bars with a lower value going to zero. The nearest distance of each transient type is estimated assuming ten years of IceCube observation. We show the estimate of FRB emission only on 10-1000 GeV as the FRB neutrino emission is negligible at lower energies. While we also show an estimate for the fluence of a Galactic core-collapse supernova occurring at a distance of 10 kpc, this does not reflect the true rate of Galactic supernovae, and it is unlikely that such a nearby event will occur in the next decade.} 
\end{figure*}

The expected neutrino fluence (time-integrated energy flux) of the nearest transients in ten years of observation is shown in Figure \ref{fluence_estimate}. The fluence estimates are evaluated using the mean values from the ranges in Table \ref{transient_values}. We compare these to the IceCube sensitivity, also integrated over the same energy range of interest, to show an appropriate comparison between the two quantities.

The uncertainty in the final fluence value is evaluated from the uncertainty in the observational parameters $x_i$ according to
$$(\Delta F_\nu)^2 =\sum_i\bigg(\frac{\partial F_\nu}{\partial x_i}\Delta x_i\bigg)^2,$$
as indicated by the errorbars in Figures \ref{fluence_estimate}. 
The parameters used to evaluate the uncertainty for each transient type are:  
\begin{itemize}
\item \textit{Shock-powered transients}: peak luminosity $L_{\text{pk}}$, mean velocity of ejecta $\bar{v}_{\text{ej}}$, rate $R_0$, particle acceleration efficiency $\epsilon_{\text{rel}}$, and spectral index $\alpha$
\item \textit{Gamma-Ray Bursts}: isotropic equivalent gamma-ray energy $\mathcal{E}_{\gamma, \text{iso}}$, bulk Lorentz factor $\Gamma$, nucleon loading factor $\xi_{\text{N}}$, and rate $R_0$. 
\item \textit{Fast Radio Bursts}: radio energy $\mathcal{E}_{\text{rad}}$ and duration $t_{\text{FRB}}$. 
\end{itemize}
The ranges of parameters used in evaluating the errorbars are intended to encompass the span of observational values for each transient type. We take the uncertainty of each variable to be half the width of its range. 

Since the aim of the predictions in Figure \ref{fluence_estimate} is to demonstrate the range of signal that could be expected from a nearby transient rather than to predict the average of a population, the calculation does not consider the distribution of source luminosity. The uncertainty calculation assumes that luminosity (or isotropic equivalent energy) and rate are independent parameters, though they could be coupled through a luminosity function. For certain source classes such as novae, supernovae, and GRBs, a luminosity function will not significantly alter the results. The luminosity functions of novae and supernovae generally follow a normal distribution \citep{De:2021cxh, Galactic_nova_rate, 2011MNRAS.412.1441L}, though they vary with subclasses and host galaxy types. The distributions of ${\cal{E}}_{\gamma, \rm iso}$ for HLGRBs and LLGRBs are relatively flat within the range considered in Table \ref{transient_values} \citep{2015MNRAS.447.1911P, Atteia:2017dcj}. In contrast, both LRNe and TDEs exhibit steep luminosity functions, $dN/dL \propto L^{-2.5}$ \citep{Karambelkar:2022tkx, 2018ApJ...852...72V} suggesting that the neutrino fluence of a future event will likely be closer to the lower end of our estimates. Finally, in the case of Galactic FRBs, both the rate and luminosity function remain largely unknown, so the effects of considering a luminosity function cannot be determined with confidence.

\section{Discussion}

The predicted fluence of all shockpowered transients falls far below the sensitivity of the IceCube Upgrade and DeepCore, indicating that sources of this nature are unlikely to be detected by future observation unless the source rates are much higher than expected. Similarly, while a Galactic FRB may have significant neutrino emission at $>$ 1 TeV, the FRB emission in GeV neutrinos falls far below the DeepCore and Upgrade sensitivity. It is worth noting that the FRB estimate used in this work is already very optimistic as it uses observational parameters consistent with the extragalactic FRB population, but is based on the distance of the nearby Galactic FRB 200428. In particular, while this event was similar to the extragalactic FRB population, it still had significantly lower energy than its extragalactic counterparts.

Notably, our analysis suggests that even Galactic novae are unlikely to be detected within the next decade. In particular, the nearby nova T Coronae Borealis \citep{TCorona, 2009ApJ...697..721S}, which is predicted to occur in the near future and has been anticipated to be a bright neutrino source, is not likely to easily be observed using IceCube DeepCore. This said, a recurrent nova may require consideration of different neutrino spectra and light curves than those used in our model, and a more precise analysis could yield more favorable results. Similarly, while the prospects of observing a nearby Galactic supernova could be favorable, with a rate of $\sim$ 3.2  per century such an event has only about a 30\% chance of occurring in the next decade.

To date, no GRB has been detected by IceCube despite dedicated GRECO searches, most notably a search for the ``brightest of all time" GRB 221009A \citep{GRECO_BOAT, GRECO_GRB}. 
The discrepancy between models predicting GRB detection the current nondetection of any GRB may be due to large uncertainties in the composition of the GRB jet. In particular, while it is often assumed that GRB jets contain a substantial population of free neutrons, this assumption is likely too optimistic. If continued observation and dedicated GRB searches yield further nondetection, this would be a strong indication that the current models used to describe GRBs are not consistent with observation, and GRBs may not harbor conditions necessary for substantial neutrino production to occur.

We note that the fluence predictions and sensitivity estimates used in this work are intended to be as broad as possible, but should not be considered to be precise. In particular, the estimates of IceCube Upgrade and DeepCore sensitivities are based on a range of spectral shapes following a power law. While this is an acceptable approximation for very short timescales, it is still not a perfect comparison to the true spectral shape of a unique transient source. Future searches, especially dedicated searches for Galactic novae or high-luminosity GRBs, could improve upon this work by tailoring the spectral shape to the shape of the particular transient in question. In addition, after the IceCube Upgrade strings are installed, a new data sample using the full upgraded
infill array (original Deep Core plus new Upgrade strings) can be developed, potentially providing enhanced sensitivity over this energy range.

\begin{table*}
    \begin{ruledtabular}
    \begin{tabular}{ccccc}
    \textbf{Source} & $\mathcal{R}_0$ (Gpc$^{-3}$/yr) & $\log_{10} L_{\mathrm{pk}}$ (erg / s) & $t_{\mathrm{pk}}$ (days) & $\bar{v}_{\mathrm{ej}}$ ($10^3$ km/s)\\  
    \hline
    \hline
    Novae & $(1 - 5) \times 10^8$ & 37 - 39 & 3 & 0.5 - 3\\
    \hline
    LRNe & $10^{5.5} - 10^{6.4}$ & 38 - 41 & 100 
    & 0.2 - 0.5\\
    \hline
    SLSNe I & $10 - 100$ & 43.3 - 44.5 & 40 
    & 5 - 10\\
    \hline
    SLSNe II & $70 - 300$ & 43.6 - 44.5 & 33.5 
    & 5 - 10\\
    \hline
    SNeIIn & 3000 & 42 - 43.7 & 35 & 5\\
    \hline
    CCSNe & $7\times10^4$ & 41.9 - 42.9 & 13.5 & 3\\
    \hline
    Type-Ia CSM & $300 - 3000$ & $\sim 43$ & 20 & 10\\
    \hline 
    TDE & $100 - 1000$ & 44 - 45 & 120 & 5 - 15\\
    \hline
    FBOT & $\sim 4800 - 8000$ & $\sim 43$ & 8 
    & 6 - 30\\
    \hline
    Lum. FBOT & $\sim 700 - 1400$ & $\sim 44$ & 3 
    & 6 - 30\\
    \hline
    \hline
    Galactic novae & 27-81 / yr & 37 - 39 & 3 & 0.5 - 3\\
    \hline
    Galactic LRNe & 0.2 / yr & 38 - 41 & 100
    & 0.2 - 0.5\\
    \hline
    Galactic CCSNe & 10 kpc & 41.9 - 42.9 & 13.5 & 3\\
    \end{tabular}
    \begin{tabular}{ccccc}
    GRB& $\mathcal{R}_0$ (Gpc$^{-3}$/yr) & $\Gamma$ & $\mathcal{E}_{\gamma, \mathrm{iso}}$ (erg) & $\xi_{N}$ \\
    \hline
    HL &  0.5 - 2 & 100 - 1000 & $10^{52} - 10^{54}$ & 3-30\\
    \hline
    LL &  200 - 500 & 10 - 100 & $10^{49} - 10^{51}$ & 3-30 \\
    \end{tabular}
    \begin{tabular}{cccc}
     \hspace{0.1 cm}& d (kpc) & $t_{\mathrm{FRB}}$ (ms) & $\mathcal{E}_{\mathrm{radio}}$ (erg)\\
     \hline
      FRB & 10 & 0.1 - 10 & $10^{37} - 10^{41}$\\
    \end{tabular}
    \end{ruledtabular}
    \caption{Ranges of values used to evaluate the estimates of Figure \ref{fluence_estimate}. The ranges of values are chosen to encompass the broad range of parameters expected for each transient type. In the case of shock-powered transients, we use observational values summarized in \citet{shock_powered_transients}; we also use range of spectral indices $\alpha = 2 - 2.7$ and particle acceleration efficiency $\epsilon_{\text{rel}} = 0.01 - 0.1$. 
    For the FRB model, we use typical FRB ranges derived by \citet{FRB_values}, and we use the distance of the Galactic event FRB 200428. For GRBs, we use ranges of bulk Lorentz factor $\Gamma$ and isotropic equivalent gamma-ray energy $\mathcal{E}_{\gamma, \text{iso}}$ consistent with ranges cited by \citet{Atteia_Egamma_iso, Murase_BOAT, GRB_decoupling} for high-luminosity and \citet{Murase_LL_GRB, Carpio_LL_GRB} for low-luminosity. We use ranges of GRB rates estimated by \citet{Murase_LL_GRB} and \citet{Swift_GRB_rate} and range of particle acceleration efficiencies presented by \citet{Murase_BOAT}.}
    \label{transient_values}
\end{table*}

\begin{acknowledgements}
The work of A.S. and K.F. is supported by the Office of the Vice Chancellor for Research and Graduate Education at the University of Wisconsin-Madison with funding from the Wisconsin Alumni Research Foundation. K.F. acknowledges support from National Science Foundation (PHY-2110821, PHY-2238916) and NASA (NMH211ZDA001N-{\it Fermi}). This work was supported by a grant from the Simons Foundation (00001470, KF). K.F acknowledges the support of the Sloan Research Fellowship.  B.D.M.~was supported in part by the NASA Fermi Guest Investigator Program through grant 80NSSC24K0408 and through the NASA ATP Program through grant 80NSSC22K080.  The Flatiron Institute is supported by the Simons Foundation.  The work of J. V. and J. T. is supported in part by Vilas Associate funding from the the Office of the Vice Chancellor for Research and Graduate Education at the University of Wisconsin-Madison and the National Science Foundation (PHY-1913607).
\end{acknowledgements}

\clearpage


\end{document}